\documentclass[aps,preprint,showpacs,preprintnumbers,amsmath,amssymb]{revtex4}
\usepackage{epsfig}
\pdfoutput=1
\usepackage{latexsym}
\usepackage{dcolumn}
\usepackage{bm}
\begin{document}
\date{\today}
\title{Algorithm to estimate the Hurst exponent\\ of
high-dimensional fractals}
\author{Anna Carbone}
 \affiliation{
 Physics Department, Politecnico di Torino, Corso Duca degli
Abruzzi 24, I-10129 Torino, Italy \\ }

\begin{abstract}
We propose an algorithm to estimate the Hurst exponent of
high-dimensional fractals, based on a generalized high-dimensional
variance around a moving average low-pass filter.  As working
examples, we consider rough surfaces generated by the Random
Midpoint Displacement  and by the Cholesky-Levinson Factorization
algorithms. The surrogate surfaces have Hurst exponents ranging
from $0.1$ to $0.9$ with step $0.1$, and different sizes. The
computational efficiency and the accuracy of the algorithm are
also discussed.

\end{abstract}
  \maketitle
\section{Introduction}   The scaling properties of random curves
and surfaces can be quantified in terms of the Hurst exponent $H$,
a parameter defined in the framework of the fractional Brownian
walks introduced in \cite{Mandelbrot}.
 A  fractional Brownian function
 $f(\bm{r}):\mathbb
{R}^d\rightarrow \mathbb{R}$, is characterized by a variance
$\sigma_H^2$:
\begin{equation}
\sigma_H^2= \,\, \left<[f({\bm{r}}+ \bm{\lambda})-f(\bm{r})]^2
\right>\,\,\propto \|\bm{\lambda}\|^{\alpha} \hspace*{15 pt} {\rm
with}\hspace*{10 pt} \alpha=2H \hspace*{3 pt}, \label{variance}
\end{equation}

\noindent with $ \bm{r}=(x_1, x_2,...,x_d)$ , $\bm{\lambda}=
(\lambda_1,\lambda_2,...,\lambda_d)$ and
$\|\bm{\lambda}\|=\sqrt{\lambda_1^2+\lambda_2^2+...+\lambda_d^2}\,$
;
 a
 power spectrum $S_H$:

\begin{equation}
S_H \propto  \|\bm{\omega}\|^{-\beta} \hspace*{25 pt} {\rm
with}\hspace*{15 pt} \beta=d+2H \hspace*{3 pt}, \label{spectrum}
\end{equation}

\noindent with $\bm{\omega} = (\omega_1,\omega_2,...,\omega_d)$
the angular frequency,
$\|\bm{\omega}\|=\sqrt{\omega_1^2+\omega_2^2+...+\omega_d^2}\,$;
 a number of objects $N_H$ of characteristic
size $\epsilon$ needed to cover the fractal:
\begin{equation}
\label{box} N_H\propto \epsilon^{-D} \hspace*{25 pt}{\rm
with}\hspace*{15 pt} D=d+1-H \hspace*{3 pt},
\end{equation}

\noindent $D$  being the fractal dimension  of $f({\bm{r}})$. The
Hurst exponent ranges from $0$ to $1$, taking the values $H=0.5$,
$H>0.5$ and $H<0.5$ respectively for uncorrelated, correlated  and
anticorrelated Brownian functions.
\par
The application of fractal concepts, through the estimate of $H$,
has been proven useful in a variety of fields. For example in
$d=1$, heartbeat intervals of healthy and sick hearts are
discriminated on the basis of the value of $H$
\cite{Thurner,Goldberger}; the stage of financial market
development is related to
 the correlation degree of return and
volatility series \cite{Dimatteo}; coding and non coding regions
of genomic sequences have different correlation degrees
\cite{Peng}; climate models are validated by analyzing long-term
correlation in atmospheric and oceanographic series
\cite{Ashkenazy,Huybers}. In $d\geq 2$ fractal measures are used
to model and quantify  stress induced morphological transformation
 \cite{Blair};  isotropic and anisotropic fracture surfaces
\cite{Ponson,Hansen,Bouchbinder,Schmittbuhl,Santucci}; static
friction between materials dominated by hard core interactions
\cite{Sokoloff}; diffusion \cite{Levitz,Malek} and transport
\cite{Oskoee,Filoche} in porous and composite materials; mass
fractal features in wet/dried gels \cite{Vollet} and in
physiological organs (e.g. lung) \cite{Suki}; hydrophobicity of
surfaces with hierarchic structure undergoing natural selection
mechanism \cite{Yang} and solubility of nanoparticles
\cite{Mihranyan}; digital elevation models \cite{Fisher} and slope
fits of planetary surfaces \cite{Sultan-Salem}.
\par  A number of fractal quantification methods - based on  the Eqs.~(\ref{variance}-\ref{box}) or
on variants of these relationships - like Rescaled Range Analysis
(R/S), Detrended Fluctuation Analysis (DFA), Detrending Moving
Average Analysis (DMA), Spectral Analysis,  have been thus
proposed to accomplish accurate and fast estimates of $H$ in order
to investigate correlations at different scales in $d=1$. A
comparatively small number of methods able to capture spatial
correlations-operating in $d \geq 2$-have been proposed so far
\cite{Rangarajan,Davies,Alvarez,Gu,Kestener,Alessio,Carbone,Arianos}.
This work is addressed to develop an
 algorithm  to estimate the Hurst exponent of
high-dimensional fractals and thus is intended to capture  scaling
 and correlation properties over space.  The proposed
method is based on a generalized high-dimensional variance of the
fractional Brownian function around a moving average.  In Section
\ref{method}, we report the relationships holding for fractals
with arbitrary dimension. It is argued that the implementation can
be carried out in directed or isotropic mode. We show that the
Detrending Moving Average (DMA) method
\cite{Alessio,Carbone,Arianos} is recovered for $d=1$. In Section
\ref{results}, the feasibility of the technique is proven by
implementing the algorithm on rough surfaces - with different size
$N_1 \times N_2$ and Hurst exponent $H$ - generated by the Random
Midpoint Displacement (RMD) and  by the Cholesky-Levinson
Factorization (CLF) methods \cite{Voss,Zhou}. The generalized
variance is estimated over sub-arrays $n_1 \times n_2$ with
different size (\emph{``scales"}) and then averaged over the whole
fractal domain $N_1 \times N_2$. This feature  reduces the bias
effects due to nonstationarity  with an overall increase of
accuracy - compared to the two-point correlation function, whose
average is calculated over all the fractal. Furthermore - compared
to the two-point correlation function, whose implementation is
carried out along $1$-dimensional lines (e.g. for the fracture
problem, the two-point correlation functions are measured along
the crack propagation direction and the perpendicular one), the
present technique is carried out over $d$-dimensional structures
(e.g. squares in $d=2$). In Section \ref{discussion}, we discuss
accuracy and range of applicability of the method.

\section{Method}
\label{method}

\noindent In order to implement the algorithm,  the generalized
variance  $\sigma^2_{DMA}$ is introduced:

\begin{equation}
 \sigma^2_{DMA}=\frac{1}{\mathcal{N}}\sum_{i_1=n_1 -m_1}^{N_1-
 m_1}\sum_{i_2=n_2 -m_2 }^{
 N_2-m_2}...\sum_{i_d=n_d -m_d }^{
 N_d-m_d }\Big[f(i_1,i_2,...,i_d)-\widetilde{f}_{n_1,n_2,...n_d}(i_1,i_2,...,i_d)\Big]^2  \hspace*{3 pt},\label{DMAd}
\end{equation}

\noindent  where $f(i_1,i_2,...,i_d)=f(\bm{i})$  is a fractional
Brownian function defined over a discrete $d$-dimensional domain,
with maximum sizes $N_1, N_2,...,\,N_d$. It is
 $i_1=1,2,...,\,N_1$, $i_2=1,2,...,\,N_2$, $...$,
$i_d=1,2,...,\,N_d$. \,  $\bm{n}=(n_1,n_2,...,\,n_d)$ defines the
sub-arrays ${\nu}_d$ of the fractal domain with maximum values
$n_{1max}=\mathrm{max}\{n_1\}$,
\,$n_{2max}=\mathrm{max}\{n_2\},...$,\,
$n_{dmax}=\mathrm{max}\{n_d\}$; $m_1=\mathrm{int}(n_1
\theta_1)$,\,$m_2=\mathrm{int}(n_2
\theta_2)$,...,\,$m_d=\mathrm{int}(n_d \theta_d)$ and $\theta_1$,
$\theta_2$,\, $...\,\theta_d$ are parameters ranging from 0 to 1;
 ${\mathcal{N}}={(N_1-n_{1max})\cdot(N_2-n_{2max})\cdot...\cdot(N_d-n_{dmax})}$.
 The function $\widetilde{f}_{n_1,n_2,...,\,n_d}(i_1,i_2,...,i_d)=\widetilde{f}$ is given by:

\begin{eqnarray}
\label{MAd} \widetilde
f_{n_1,n_2,...,\,n_d}(i_1,i_2,...,\,i_d)=\frac{1}{n_1 n_2 ...
n_d}  \sum_{k_1=-m_1}^{n_1-1-m_1} \sum_{k_2=-m_2}^{n_2-1-m_2}...
\nonumber
\\ ...\sum_{k_d=-m_d}^{n_d-1-m_d} f(i_1-k_1,i_2-k_2,..., i_d-k_d)
\hspace*{3 pt} ,
\end{eqnarray}

\noindent that is an average of $f$ calculated over the sub-arrays
${\nu}_d$. The Eqs.~(\ref{DMAd}) and (\ref{MAd}) are defined for
any  value of $n_1,n_2,...,\,n_d$ and for any shape of the
sub-arrays, however, it is preferable to choose sub-arrays with
$n_1=n_2=....=n_d$ to avoid spurious directionality in the
results. The generalized variance $\sigma^2_{DMA}$  varies as
$(\sqrt{n_1^2+n_2^2+...+n_d^2}\,)^{2H}$  as a consequence of the
property (\ref{variance}) of the fractional Brownian functions.
\par
Upon variation of the parameters $\theta_1$,
$\theta_2\,,...\,,\theta_d$ in the range $\left[ 0, 1 \right]$,
the indexes $i_1,i_2,...,\,i_d$ and $k_1$, $k_2\,,...\,, k_d$ of
the sums in the Eqs.~(\ref{DMAd}) and (\ref{MAd}) are accordingly
set within  ${\nu}_d$. In particular,  $(i_1,i_2,...,\,i_d)$
coincides respectively with: (a) one of the vertices of ${\nu}_d$
for $\theta_1=\theta_2=...=\theta_d=0$ and
$\theta_1=\theta_2=...=\theta_d=1$ or (b) the center of ${\nu}_d$
for $\theta_1=\theta_2=...=\theta_d=1/2$. It is worthy of note
that the choice $\theta_1=\theta_2=...=\theta_d=1/2$  corresponds
to the \emph{isotropic} implementation of the algorithm, while
$\theta_1=\theta_2=...=\theta_d=0$ and
$\theta_1=\theta_2=...=\theta_d=1$ correspond to the
\emph{directed} implementation.  For example in $d=2$, the
\emph{isotropic} implementation implies that the variance
 defined by the Eq.~(\ref{DMAd}) is referred  to a moving average $\widetilde f$ calculated over squares $n_1 \times n_2$ whose center is
$(i_1,i_2)$. Conversely, the \emph{directed} implementation
implies that the function $\widetilde f$ is calculated over
squares $n_1 \times n_2$  with one of the vertices in $(i_1,i_2)$.
The \emph{directed} mode is of interest to estimate $H$ in
fractals with preferential growth direction,
 e.g. biological tissues (lung), epitaxial layers, crack propagation in
fracture (\emph{anisotropic fractals}). If the fractal is
\emph{isotropic} and the accuracy is a priority, the parameters
$\theta_1,\theta_2,...,\theta_d$ should be preferably taken equal
to $1/2$ to achieve the most precise estimate of $H$. The
dependence of the algorithm on $\theta$ for $d=1$ has been
discussed  in \cite{Arianos}.
\par In order to calculate the Hurst exponent, the algorithm
is implemented through the following steps. The moving average
$\widetilde{f}$ is calculated for different sub-arrays ${\nu}_d$,
by varying $n_1,n_2,...,\,n_d$ from 2 to the maximum values
$n_{1max},n_{2max},...,\,n_{dmax}$. The values
$n_{1max},n_{2max},...,\,n_{dmax}$ depend on the maximum size of
the fractal domain. In order to minimize the saturation effects
due to finite-size, it should be: $n_{1max}<<N_1$;
$n_{2max}<<N_2;...;\,n_{dmax}<<N_d$. These constraints will be
further clarified in Section \ref{results}, where the algorithm is
implemented over fractal surfaces with different sizes. For each
sub-array ${\nu}_d$, the corresponding value of $\sigma_{DMA}^2$
is calculated and finally plotted on log-log axes.

\noindent To elucidate the way the algorithm works, in the
following we consider its implementation for $d=1$ and $d=2$. The
case $d=1$ reduces to the  Detrending Moving Average (DMA) method
already used for long-range correlated time series
\cite{Alessio,Carbone,Arianos}.
\medskip

 \noindent \emph{1-dimensional case:}

\noindent By posing $d=1$ in the Eq. (\ref{DMAd}), one obtains:

\begin{equation}
 \sigma^2_{DMA}=\frac{1}{N_1- n_{1max}}\sum_{i_1=n_1-m_1}^{N_1-
 m_1}\Big[f(i_1)-\tilde{f}_{n_1}(i_1)\Big]^2  \hspace*{3 pt}, \label{DMA1}
\end{equation}

\noindent where $N_1$ is the length of the sequence, $n_1$ is the
sliding window and $n_{1max}=\mathrm{max}\{n_1\}\ll N_1$. The
quantity $m_1=\mathrm{int}(n_1 \theta_1)$ is the integer part of
$n_1 \theta_1$ and $\theta_1$ is a parameter ranging from 0 to 1.
The relationship (\ref{DMA1}) defines a generalized variance of
the sequence $f(i_1)$ with respect to the function
$\widetilde{f}_{n_1}(i_1)$:

\begin{equation}
\widetilde{f}_{n_1}(i_1)= \frac{1}{n_1}\sum_{k_1=- m_1}^{n_1-1-m_1
}f(i_1-k_1)  \hspace*{3 pt},\label{MA1}
\end{equation}

\noindent which is the moving average of $f(i_1)$ over each
sliding window of length $n_1$.   The moving average
$\widetilde{f}_{n_1}(i_1)$ is calculated for different values of
the window $n_1$, ranging from 2 to the maximum value $n_{1max}$.
The variance $\sigma_{DMA}^2$ is then calculated according to the
Eq.~(\ref{DMA1}) and plotted as a function of $n_1$ on log-log
axes. The  plot is a straight line, as expected for a power-law
dependence of $\sigma_{DMA}^2$ on $n_1$:

\begin{equation}
 \sigma_{DMA}^{2} \sim n_1^{2H} \hspace*{3 pt}.
  \label{DMAscale}
\end{equation}

 \noindent The Eq.~(\ref{DMAscale})  allows one to estimate
the scaling exponent $H$ of the series $f(i_1)$. Upon variation
of  the parameter $\theta_1$   in the range $\left[ 0, 1 \right]$,
the  index $k_1$   in $\widetilde{f}_{n_1}(i_1)$ is accordingly
set within the window $n_1$. In particular, $\theta_1=0$
corresponds to  average ${f}_{n_1}(i_1)$ over all the points to
the left of $i_1$ within the window $n_1$;
 $\theta_1=1$ corresponds to  average ${f}_{n_1}(i_1)$ over all
the points to the right of $i_1$ within the window $n_1$;
$\theta_1=\frac{1}{2}$ corresponds to average ${f}_{n_1}(i_1)$
with the reference point in the center of the window $n_1$.

\medskip

\noindent \emph{2-dimensional case}

\noindent For $d=2$, the generalized variance defined by the
Eq.(\ref{DMAd}) writes:

\begin{equation}
 \sigma^2_{DMA}=\frac{1}{(N_1-n_{1max})(N_2-n_{2max})}\sum_{i_1=n_1 -m_1 }^{N_1-
 m_1}\sum_{i_2=n_2-m_2 }^{
 N_2-m_2}\Big[f(i_1,i_2)-\widetilde{f}_{n_1,n_2}(i_1,i_2)\Big]^2  \hspace*{3 pt}, \label{DMA2}
\end{equation}

\noindent with $\widetilde{f}_{n_1,n_2}(i_1,i_2)$ given by:

\begin{equation}
\label{MA2} \widetilde f_{n_1,n_2}(i_1,i_2)=\frac{1}{n_1 n_2}
\sum_{k_1=-m_1}^{n_1-1-m_1} \sum_{k_2=-m_2}^{n_2-1-m_2}
f(i_1-k_1,i_2-k_2) \hspace*{5 pt} .
\end{equation}

\noindent The average $\widetilde{f}$ is calculated over
sub-arrays with different size $n_1 \times n_2$.
 The next
step  is the calculation of the difference $f(i_1,i_2)-\widetilde
f_{n_1,n_2}(i_1,i_2)$ for each sub-array $n_1\times n_2$. A
log-log plot of $\sigma_{DMA}^2$:

\begin{equation}
 \sigma_{DMA}^{2} \sim \left[\sqrt{n_1^2+n_2^2}\right]^{2H}\sim {s}^{H} \hspace*{5 pt} .
 \label{sigma2b}
\end{equation}

\noindent as a function of  $s=n_1^2+n_2^2$, yields a straight
line with slope $H$.

\noindent Depending upon the values of the parameters $\theta_1$
and $\theta_2$, entering the quantities $m_1=\mathrm{int}(n_1
\theta_1)$ and $m_2=\mathrm{int}(n_2 \theta_2)$ in the
Eqs.~(\ref{DMA2},\ref{MA2}), the position of  $(k_1,k_2)$ and
$(i_1,i_2)$ can be varied within each sub-array. $(i_1,i_2)$
coincides with a vertex of the sub-array if: ({\em i})
$\theta_1=0$, $\theta_2=0$; ({\em ii}) $\theta_1=0$, $\theta_2=1$;
({\em iii}) $\theta_1=1$, $\theta_2=0$; ({\em iv}) $\theta_1=1$,
$\theta_2=1$ (\emph{directed implementation}). The choice
$\theta_1=\theta_2=1/2$ corresponds to take the
 point $(i_1,i_2)$ coinciding with the center of each sub-array
$n_1 \times n_2$ (\emph{isotropic implementation}) \cite{note1}.

\medskip
\section{Results}
\label{results}
  \noindent In order to test feasibility and robustness of the proposed method,  synthetic rough surfaces with assigned Hurst exponents have been generated by
the Random Midpoint Displacement (RMD) algorithm   and by the
Cholesky-Levinson Factorization (CLF) method \cite{Voss,Zhou}. The
widespread use of the RMD  algorithm is based on the trade-off of
its fast, simple and efficient implementation to its limited
accuracy especially for $H \ll 0.5$ and $H \gg 0.5$. Conversely,
the Cholesky-Levinson Factorization method is one of the most
accurate techniques to generate $1d$ and $2d$ fractional Brownian
functions, at the expenses of a more complex implementation
structure \cite{note2}.

\begin{figure}
\includegraphics[width=8cm,height=6cm,angle=0]{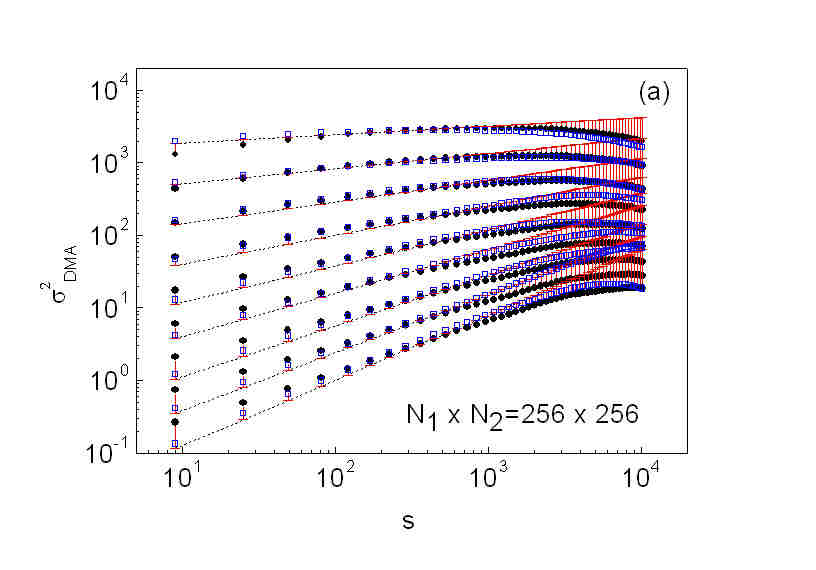}
\includegraphics[width=8cm,height=6cm,angle=0]{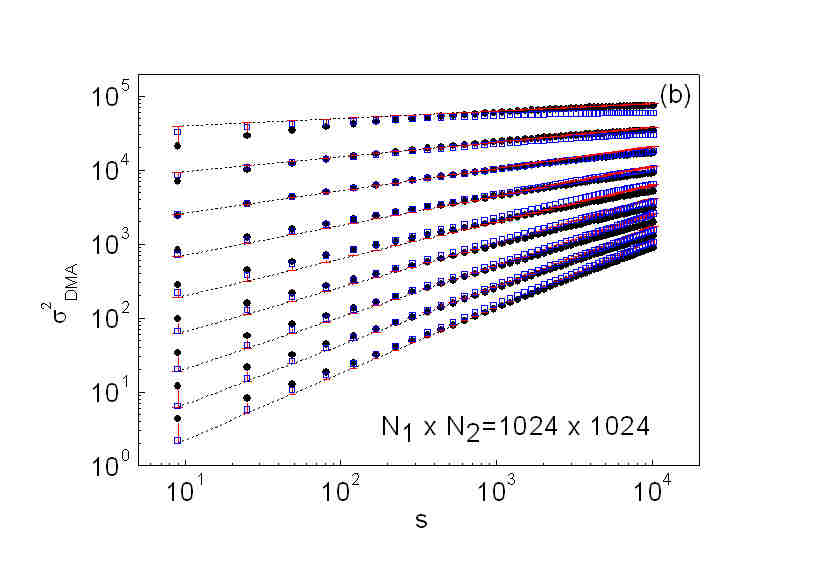}
\includegraphics[width=8cm,height=6cm,angle=0]{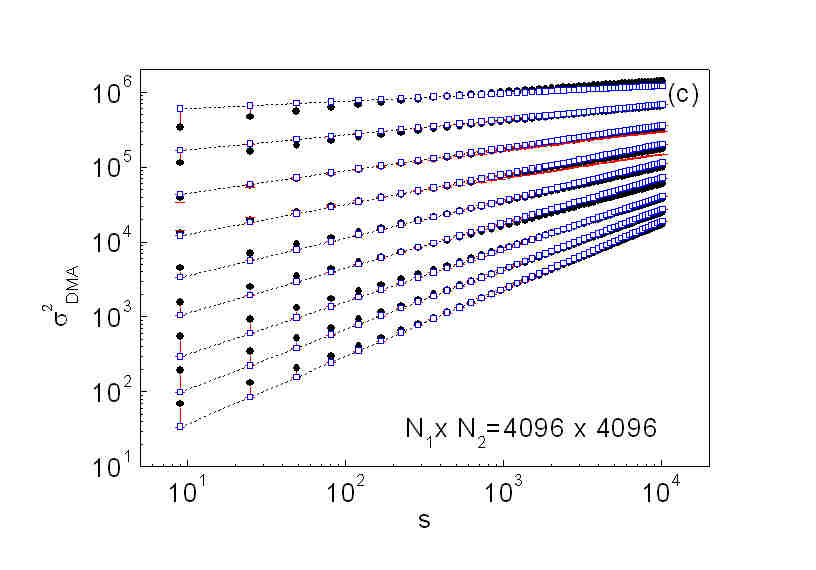}
\caption{\label{fig:DMA} (Color online)  Log-log plot of
$\sigma^2_{DMA}$ for fractal surfaces respectively with size $N_1
\times N_2= 256 \times 256 $ (a), $N_1 \times N_2=1024 \times
1024$ (b) and $N_1 \times N_2=4096 \times 4096$ (c). The data
refer to fractal surfaces generated by the RMD (circles) and by
the CLF (squares) methods. The Hurst exponent $H_{in}$ - input of
the RMD and the CLF algorithm - varies from $0.1$ to $0.9$ with
step $0.1$. The results correspond to the isotropic
implementation, i.e. with the parameters $\theta_1=\theta_2=1/2$
in the Eq.(\ref{DMA2}). The dashed lines represent the behavior
expected for full linearity, i.e. the log-log plot of curves
varying as $s^{H_{in}}$. It is worthy of note that the CLF data
are closer to the full-linearity compared to the RMD ones.}
\end{figure}

\begin{figure}
\includegraphics[width=9cm,height=6cm,angle=0]{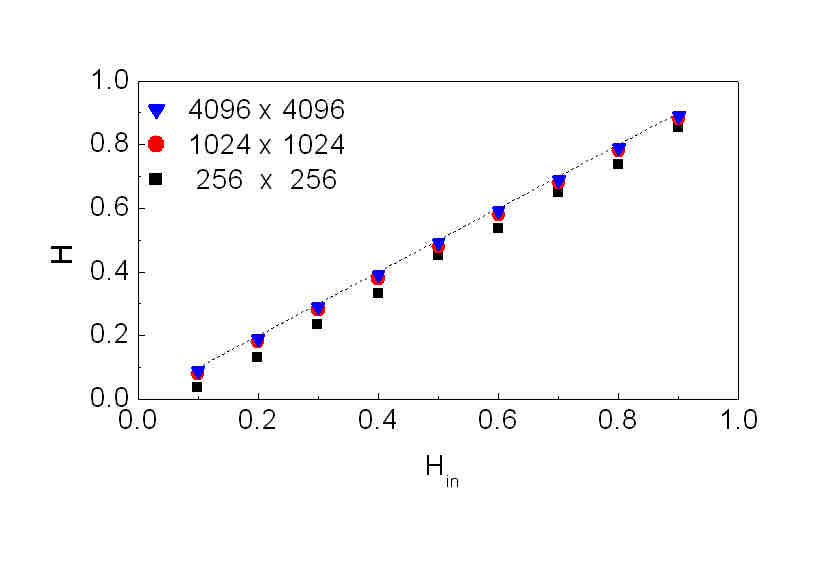}
\caption{\label{fig:all} (Color online) Plot of the values of $H$
obtained by linear fit of the curves shown in Fig. \ref{fig:DMA}
(a), (b), (c). The data refer to the fractal surfaces generated by
the Cholesky-Levinson Factorization method (squares). The dashed
lines represent the ideal behavior: $H=H_{in}$. }
\end{figure}

In Fig.~\ref{fig:DMA}, the log-log plots of $\sigma^2_{DMA}$ as a
function of $s$ are shown for the synthetic fractal surfaces
generated by the RMD (circles) and by the CLF method (squares).
 The surfaces have
Hurst exponents $H_{in}$ ranging from $0.1$ to $0.9$ with step
$0.1$. The domain sizes are respectively $N_1 \times N_2=256
\times 256 $ (a), $N_1 \times N_2=1024 \times 1024$ (b) and $N_1
\times N_2=4096 \times 4096$ (c).
 The dashed lines
show the behavior that should be exhibited by variances varying
exactly as $s^{H_{in}}$ over the entire range of scales. The plots
of $\sigma^2_{DMA}$ as a function  $s$ are in good agreement with
the behavior expected on the basis of the Eq.~(\ref{sigma2b}).
 The quality of the fits is higher for the surfaces generated by the
CLF method, confirming that the RMD algorithm synthesizes less
accurate fractals. By comparing the results of the simulation
(symbols) to the straight lines corresponding to full linearity
over the whole range (dashed), deviations from the full linearity
can be observed especially for the small surfaces at the extremes
of the scale. A plot of the slopes for the fractal surfaces
generated by the CLF algorithm is shown in Fig. \ref{fig:all} for
different sizes of the fractal domain. A detailed discussion of
the origin of the deviations at low and large scales is reported
in the Section \ref{discussion}.

\begin{figure}
      \includegraphics[width=5cm,height=4cm,angle=0]{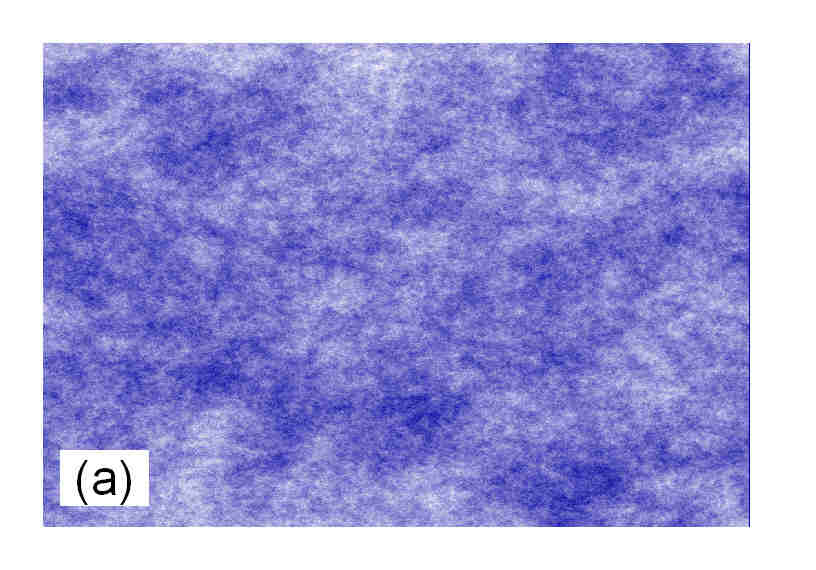}
\includegraphics[width=5cm,height=4cm,angle=0]{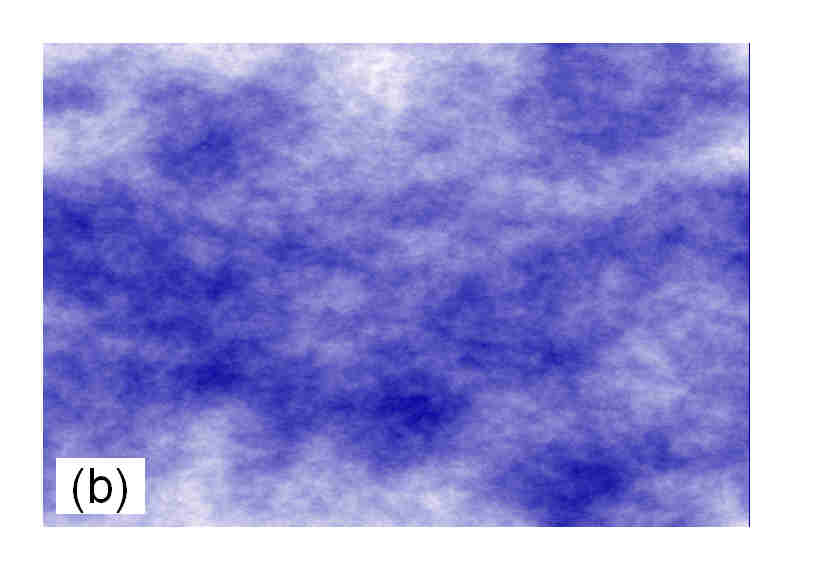}
      \includegraphics[width=5cm,height=4cm,angle=0]{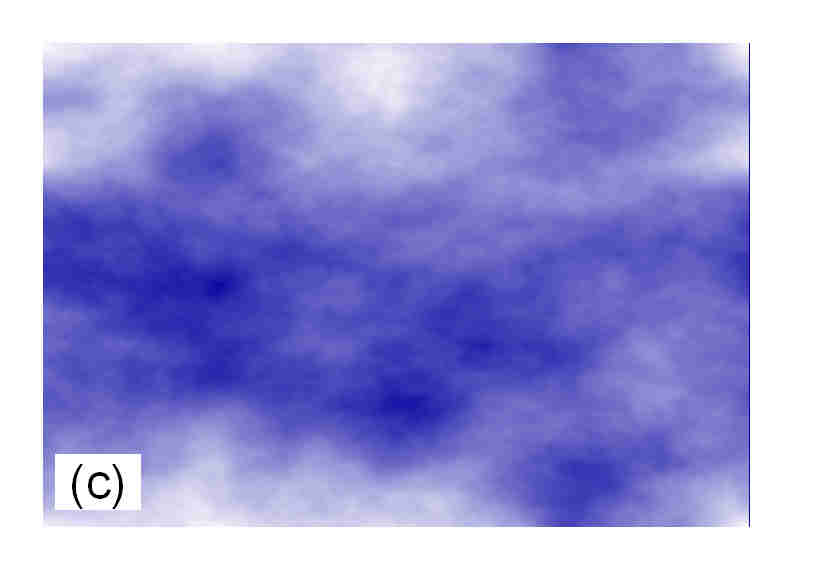}
  \caption{(Color online). Cloudy sky images respectively with Hurst
  exponent $H=0.1$ (a), $H=0.5$  (b) and $H=0.9$ (c). Such heterogeneous images are represented as fractal surfaces
  by mapping the color intensity
   (RGB
   content).}
  \label{sky}
  \end{figure}

\noindent
 Finally, we also show three examples of digital images currently
mapped to fractal surfaces  with reference to the color intensity
i.e. to the levels of  Red, Green and Blue (RGB). The Hurst
exponents estimated by the proposed method are respectively
$H=0.1$ (a), $H=0.5$ (b) and $H=0.9$ (c) for the images in
Fig.~\ref{sky}.

\section{Discussion}
\label{discussion} The proposed algorithm is characterized by
short execution time and ease of implementation. By considering
the case $d=2$, the function $\widetilde f_{n_1,n_2}(i_1,i_2)$ is
indeed simply obtained by summing the values of $f(i_1,i_2)$ over
each sub-array $n_1 \times n_2$. Then the sum is updated at each
step by adding the last and discarding the first row (column) of
each sliding array $n_1 \times n_2$. For higher dimensions, the
sum is updated at each step by adding and discarding a $d-1$
dimensional set of each array $n_1 \times n_2 \times ... \times
n_d$. The algorithm does not use
 arbitrary parameters, the computation simply relying on averages of $f$.
 We will now argue on the origin of the deviations at
 small and at large scales.

\par
\emph{Deviations at large scales. }
 The  deviations from the linearity  at large scales, leading to the
saturation of the $\sigma^2_{DMA}$,  are due to finite size
effects. The small surfaces do not contain enough data to make the
evaluation of the scaling law over the  sub-arrays statistically
meaningful.   By comparing the data in
Figs.~\ref{fig:DMA}~(a),~(b),~(c), one can note that the
saturation effect  decreases upon increasing the size $N_1 \times
N_2$ of the fractal surface. The finite size effects become
negligible when the conditions $n_{1max}<<N_1$;
$n_{2max}<<N_2;...;\,n_{dmax}<<N_d$ are fulfilled.

\par
\begin{figure}
\includegraphics[width=8cm,height=6cm,angle=0]{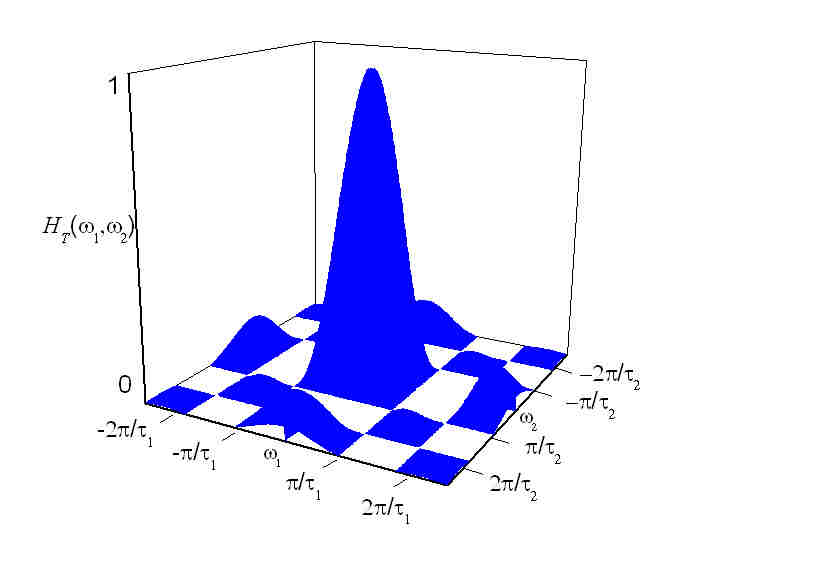}
\caption{\label{Homega} (Color online) Plot of the transfer
function $\mathcal{H}_\mathcal{T}(\omega_1,\omega_2)$.}
\end{figure}

\emph{Deviations occurring at small scales.} The deviations
occurring at low scales are related to the departure of the
low-pass filter from the ideality. This problem also occurs with
one-dimensional fractals (time series) resulting in the quite
generally reported overestimation of $H$ in anticorrelated signals
 and underestimation of $H$ in correlated
signals  \cite{Caccia,Hu,Chen,Xu,Stoev}. We will discuss the
origin of these deviations by means of the filter transfer
function $\mathcal{H}_\mathcal{T}(\omega)$ \cite{Hamming}.
 The algorithm is based on a
generalized variance of the function $f$ with respect to
$\widetilde{f}$. The function $\widetilde{f}$ is
 the output of a
\emph{low-pass filter} driven by $f$, with impulse response a
box-car function. In the Appendix, the transfer function
$\mathcal{H}_\mathcal{T}(\omega)$ of $\widetilde{f}$ is
explicitly  calculated and shown in Fig.(\ref{Homega}) for $d=2$.
 For an
\emph{ideal} low-pass filter, the transfer function  should be
one or zero respectively at frequencies lower or higher than the
cut-off frequency.
  However, in real low-pass filters, at frequencies lower than the cut-off frequency, all the
components of the signal suffer some attenuation but $\omega=0$.
The cut-off frequencies
 of  $\mathcal{H}_\mathcal{T}(\omega)$ are
$\omega_i=\pi/\tau_i$, i.e. the first zeroes of the functions
${\sin \omega_{i}\tau_{i}}/{\omega_{i} \tau_{i}}$ in the
Eq.~(\ref{Transfer function2}).
   Moreover, in real filters, at frequencies higher than $\pi/\tau_i$,  due to the presence of the
sidelobes, components of the signals lying in the bands
$(\pi/\tau_{i}, 2 \pi/\tau_{i});
(2\pi/\tau_{i},
3\pi/\tau_{i});...$, are not fully filtered out.
    As a result, the function $\widetilde{f}$ contains: (a)
  less components with frequency lower than $\omega_i=\pi/\tau_i$ and (b)
more components with frequency higher than $\omega_i=\pi/\tau_i$
compared to what it would be expected with an ideal low-pass
filter. The lack of low-frequency components depends on the
central lobe, while the excess of high-frequency components
depends on the side lobes. The excess of high-frequency components
results in a smaller value of the difference $f-\widetilde f$,
i.e. in a decrease of $\sigma_{DMA}^2$ and, thus, in an increase
 of the slope of the
log-log plot. Conversely, the lack of low-frequency components
results in a larger value of the difference $f-\widetilde f$, i.e.
in an increase of $\sigma_{DMA}^2$ and, thus, in a decrease
 of the slope of the
log-log plot. The two
 effects are more relevant with smaller values of the scales, when the filter nonideality is
 greater. Moreover, as one can deduce from the
Eqs.~(\ref{spectrum}) and (\ref{Transfer Spectrum}), the effect of
the side lobes dominates in high-frequency  rich fractals with $H
< 0.5$, while the effect of the central lobe is dominant in
fractals with $H> 0.5$, rich of low-frequency components.

\begin{table*}
\caption{\label{tab:table} Slopes $H_{I}$, $H_{II}$, $H_{III}$ and
relative errors $\Delta H_{I}$, $\Delta H_{II}$, $\Delta H_{III}$
of the curves plotted in Fig.~\ref{fig:DMA}(b) (squares). The
slopes have been calculated by linear fit respectively over the
ranges: $10\leq s \leq 100$ ($H_{I}$), $10\leq s \leq 1000$
($H_{II}$) and $10\leq s \leq 10000$ ($H_{III}$). The errors
$\Delta H_{I}$, $\Delta H_{II}$, $\Delta H_{III}$ are calculated
as $\Delta H=(H-H_{in})/H_{in}$. }
\begin{ruledtabular}
\begin{tabular}{ccccccc}
 $H_{in}$& $H_{I}$ & $\Delta H_{I}$&  $H_{II}$ & $\Delta H_{II}$&  $H_{III}$ &
 $\Delta H_{III}$
\\ \hline
 0.1&$0.1346$& $+3.46 \cdot 10^{-1}$&$0.1073$ &$+7.30 \cdot 10^{-2}$& 0.0718 &$-2.822 \cdot 10^{-1}$  \\
 0.2&$0.2272$& $+1.36 \cdot 10^{-1}$&$0.2050$ &$+2.50 \cdot 10^{-2}$& 0.1700  &$-1.500 \cdot 10^{-1}$  \\
 0.3&$0.3233$& $+7.77 \cdot 10^{-2}$&$0.2995$ &$-1.67 \cdot 10^{-3}$& 0.2716 &$-9.467 \cdot 10^{-2}$   \\
 0.4&$0.4205$& $+5.12 \cdot 10^{-2}$&$0.3970$ &$-7.50 \cdot 10^{-3}$& 0.3691  &$-7.725 \cdot 10^{-2}$  \\
 0.5&$0.5178$& $+3.56 \cdot 10^{-2}$&$0.4973$ &$-5.40 \cdot 10^{-3}$& 0.4752  & $-4.960 \cdot 10^{-2}$ \\
 0.6&$0.6171$& $+2.85 \cdot 10^{-2}$&$0.5973$ &$-4.50 \cdot 10^{-3}$& 0.5617  & $-6.383 \cdot 10^{-2}$  \\
 0.7&$0.7185$& $+2.64 \cdot 10^{-2}$&$0.6956$ &$-6.29 \cdot 10^{-3}$& 0.6770  & $-3.286 \cdot 10^{-2}$  \\
 0.8&$0.8207$& $+2.58 \cdot 10^{-2}$&$0.7999$ &$-1.25 \cdot 10^{-4}$& 0.7659 & $-4.263 \cdot 10^{-2}$  \\
 0.9&$0.9253$& $+2.81 \cdot 10^{-2}$&$0.8999$ &$-1.11 \cdot 10^{-4}$& 0.8679  & $-3.567 \cdot 10^{-2}$ \\
\end{tabular}
\end{ruledtabular}
\end{table*}
\par   In order to gain further insight in the above theoretical arguments, we
report in Table \ref{tab:table} the slopes $H_I$, $H_{II}$ and
$H_{III}$  of the curves (squares) plotted in
Fig.~\ref{fig:DMA}~(b))  over different ranges. The slopes have
been calculated by linear fit respectively over the ranges $10\leq
s \leq 100$ ($H_I$), $10\leq s \leq 1000$ ($H_{II}$) and $10\leq s
\leq 10000$ ($H_{III}$). The relative errors $\Delta
H=(H-H_{in})/H_{in}$ are given respectively in the $3^{rd}$,
$5^{th}$ and $7^{th}$ columns. The slope $H_I$ is greater than the
expected value $H_{in}$. The slope $H_{II}$ is overestimated for
$H=0.1$ and $H=0.2$ and underestimated for $H>0.2$. The slope
$H_{III}$ is underestimated since the effects of the finite-size
of the fractal domain play a dominant role.
\par
\begin{figure}
\includegraphics[width=8cm,height=6cm,angle=0]{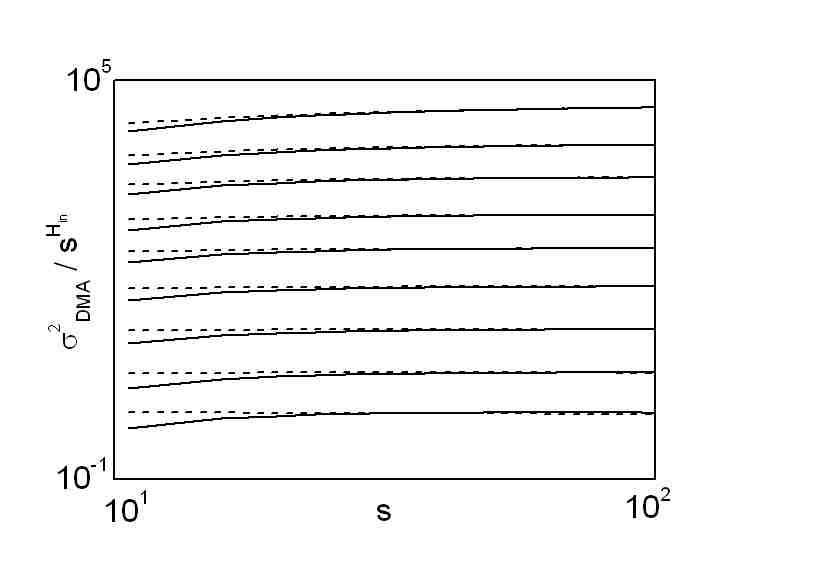}
\caption{\label{wma} (Color online) Plot of the function
$\sigma^2_{DMA}$  with $\widetilde f$ defined by the
Eq.~(\ref{DMAd})  (solid lines) and $\widetilde f^\star$  defined
by the Eq.~(\ref{WMA}) (dashed lines). It can be noted that the
deviations of the slope at small scales are reduced by the use of
$\widetilde f^\star$ implying a corresponding reduction of the
relative error $\Delta H_{I}$ of two orders of magnitude.}
\end{figure}

We address the question if the artifacts due to the filter
nonideality described above might be  corrected somehow. In the
remaining of this section, we will thus consider the use of
\emph{{windows}} whose general effect is to increase the width of
the central lobe while reducing those of the sidelobes of the
function $\mathcal{H}_\mathcal{T}(\omega)$ (a detailed description
of these methods can be found in \cite{Hamming}). By restricting
our discussion to the present technique, the correction is
performed by using the following variant of the relationship
(\ref{MAd}):

\begin{eqnarray}
\label{WMA} \widetilde f^\star_
{n_1,n_2,...,\,n_d}(i_1,i_2,...,\,i_d) =&&(1-\alpha)
f_{n_1,n_2,...,\,n_d}(i_1,i_2,...,\,i_d) \nonumber\\ && + \alpha
\widetilde f_{n_1,n_2,...,\,n_d}(i_1-1,i_2-1,...,\,i_d-1)
\hspace*{3 pt} ,
\end{eqnarray}
\noindent where $\alpha=n_1n_2...n_d/[(n_1+1)(n_2+1)...(n_d+1)]$.
The Eq.~(\ref{WMA})  reduces for $d=1$ to the exponentially
weighted moving average (EWMA). In practice,  the difference
between the Eq.~(\ref{MAd}) and the Eq.~(\ref{WMA}) is that the
function $\widetilde f^\star$ places more importance to the data
around the point $i_1,i_2,...,\,i_d$. This is achieved  by
assigning to the function a weight $(1-\alpha)$, while all the
other values are summed together and weighted by $\alpha$. In
Fig.~\ref{wma}, we show the ratio $\sigma^2_{DMA}/s^{H_{in}}$
obtained by implementing the algorithm respectively with the
function $\widetilde f$ (solid lines) and $\widetilde f^\star$
(dashed lines) in the range $10\leq s\leq 100$. The ratio
$\sigma^2_{DMA}/s^{H_{in}}$ is noticeably closer to a constant
value when the function $\widetilde f$ is replaced by $\widetilde
f^\star$, with a corresponding reduction of two orders of
magnitude in the relative error $\Delta H_{I}$.

\section{Conclusion}

We have put forward an algorithm to estimate the Hurst exponent of
fractals with arbitrary dimension, based on the high-dimensional
generalized variance $\sigma^2_{DMA}$ defined by the Eq.
(\ref{DMAd}).
\par
The methods currently used to estimate the Hurst exponent of
high-dimensional fractals are based on: (\emph{i})  $1-d$
two-point correlation and structure functions operated along
different directions, (\emph{ii}) high$-d$ Fourier and wavelet
transforms
\cite{Ponson,Hansen,Bouchbinder,Schmittbuhl,Santucci,Kestener}.
 The
advantage of the methods (\emph{i}) is the ease of implementation.
Their drawback is the limited accuracy due to biases and
nonstationarities, being these functions calculated over the
entire fractal domain. The methods (\emph{ii})  are more accurate,
however their implementation is complicated especially for data
set with limited extension. The generalized variance
$\sigma_{DMA}^2$ is \emph{``scaled"}, meaning that it is
calculated over sub-arrays of the whole fractal domain by means of
the function $\widetilde f$. The \emph{``scales"} are set by the
size of the sub-arrays $n_1 \times n_2 \times.... \times n_d$.
Therefore, the proposed method exhibits at the same time: (a) ease
of implementation, being based on a variance-like approach and (b)
high accuracy, being
 calculated over scaled sub-arrays rather
than on the whole fractal domain.

A further important feature of the proposed algorithm is that it
can be implemented \emph{``isotropically"} or in
\emph{``directed"} mode to accomplish estimates of $H$ in fractals
having preferential growth direction e.g. biological tissues,
epitaxial layers or in crack propagation in fracture. The
\emph{isotropic implementation} is obtained by taking $\theta_1=
\theta_2=...=\theta_d=1/2$
 in the Eq.~(\ref{DMAd}). This choice implies that the reference
point  $(i_1,i_2,...,i_d)$ of the moving average lies in the
center of each sub-array $n_1\times n_2 \times ...\times n_d$ and
thus $\widetilde f$ is calculated by summing the values of $f$
around $(i_1,i_2,...,i_d)$. Conversely, to implement the algorithm
in a preferential direction (\emph{directed implementation}), the
reference point must be coincident with one of extremes of the
segment $n_1$, or with one of the vertices of the square grid
$n_1\times n_2$ or of the d-dimensional array $n_1\times n_2
\times ...\times n_d$. The directed implementation can be
performed by choosing for example $\theta_1=
\theta_2=...=\theta_d=0$. \par Further generalizations of the
proposed method can be envisaged for applications to the analysis
of time-dependent spatial correlations in $d \geq 2$.

\appendix*
\section{Transfer Function of $\widetilde f$ }

The function $\widetilde f$, defined by the Eq.~(\ref{MAd}),
corresponds  to the discrete form of the integral:
\begin{eqnarray}\label{ConvolutiondD}
\widetilde f(x_1, x_2, ...,\, x_d)=\frac{1}{\tau_1 \tau_2...
\tau_d}\int_{x_1-\tau_1}^{x_{1}} \! dx_1'\nonumber\\
\int_{x_2-\tau_2}^{x_{2}}
\!dx_2'\,...\int_{x_d-\tau_d}^{x_{d}}dx_d' f(x_1',x_2',...,x_d')
\end{eqnarray}
\noindent where for the sake of simplicity we have considered the
case $\theta_1=\theta_2=,...,=\theta_d=0$.

\noindent The Eq.~(\ref{ConvolutiondD}) can be rewritten as a
convolution integral:
$$
\widetilde f(x_1, x_2,...,\,x_d)=\frac{1}{\tau_1 \tau_2...
\tau_d}\int_{-\infty}^\infty dx_1^*\,U \!\!
\left(\frac{x_1^*}{{\tau_1}}\right)\int_{-\infty}^\infty dx_2^*
\,U \!\!\left(\frac{x_2^*}{\tau_2}\right) ...
$$
\begin{equation}
\label{ConvolutiondD3} ...\int_{-\infty}^\infty dx_d^* \,U
\!\!\left(\frac {x_d^*}{{\tau_d}}\right)
f(x_1-x_1^*,x_2-x_2^*,...,\,x_d-x_d^*)
\end{equation}

\noindent  with the convolution kernels  given by the boxcar
function:

$$
  U(x_i^*/\tau_i)=
\begin{cases} &  1 \hspace*{5 pt}  {\rm for}  \hspace*{5 pt}  0<x^*/\tau_i<1 \\
  &  0  \hspace*{5 pt}  {\rm elsewhere} \hspace*{5 pt} .\\
   \end{cases}
    \label{boxcar}
$$
\noindent The transfer function can be calculated as follows:

\begin{eqnarray}\label{Transfer function1}
   \mathcal{H}_\mathcal{T}(\omega_1,\omega_2,...\, ,\omega_d)= \frac{1}{\tau_1\tau_2
   ...
    \tau_d}  \int_{0}^{\tau_1} dx_1\int_{0}^{\tau_2} dx_2\,...\nonumber\\\int_{0}^{\tau_d}
     dx_d \exp [-i2\pi (\omega_1 x_1+\omega_2
    x_2+...+\omega_d x_d)]
\end{eqnarray}

\noindent that can be written as:

\begin{equation}\label{Transfer function2}
    \mathcal{H}_\mathcal{T}(\omega_1,\omega_2,...,\omega_d)= \prod_{i=1}^d \frac{\sin
    \omega_{i}
    \tau_{i}}{\omega_{i}\tau_{i}}
\end{equation}
\noindent that is thus $d$-times the  function ${\sin
\omega_{i}\tau_{i}}/{\omega_{i} \tau_{i}}$. \par The Fourier
transform $\widetilde{ \mathcal{F}}$ of the function $\widetilde
f$  can be obtained by means of the following relationship:

\begin{equation}\label{Transfer function}
  \widetilde{ \mathcal{F}}(\omega_1,\omega_2,...,\,\omega_d)= \mathcal{H}_\mathcal{T}(\omega_1,\omega_2,...,\,\omega_d) \mathcal{F}(\omega_1,\omega_2,...\omega_d)
\end{equation}
\noindent where $\mathcal{F}(\omega_1,\omega_2,...,\omega_d)$ is
the Fourier transform of the function $f(x_1,x_2,...,\,x_d)$.
\par \noindent The power spectrum ${\widetilde S}$ of the
function $\widetilde f$ is given by:

\begin{equation}\label{Transfer Spectrum}
 {\widetilde S} (\omega_1,\omega_2,...,\omega_d)= |\mathcal{H}_\mathcal{T}(\omega_1,\omega_2,...,\omega_d)|^2 S(\omega_1,\omega_2,...,\omega_d)
\end{equation}

\noindent where $S(\omega_1,\omega_2,...,\omega_d)$ is the power
spectrum of the function $f(x_1,x_2,...,x_d)$.
\par \noindent

%
%

\end{document}